\documentclass[11pt,sort&compress]{elsarticle}
\makeatletter
\def\ps@pprintTitle{%
 \let\@oddhead\@empty
 \let\@evenhead\@empty
 \def\@oddfoot{\centerline{\thepage}}%
 \let\@evenfoot\@oddfoot}
\makeatother
\usepackage{amsmath}
\usepackage{accents}
\usepackage{amssymb}
\usepackage{mathrsfs}
\usepackage[LGR,T1]{fontenc}
\usepackage[latin1]{inputenc}
\usepackage[cal=boondox,calscaled=1]{mathalfa}

\DeclareMathAlphabet{\bbvar}{U}{BOONDOX-ds}{m}{n}
\DeclareMathAlphabet{\bbgreek}{U}{bbold}{m}{n}
\usepackage[defaultsans,osfigures,scale=0.9]{opensans}
\usepackage{psfrag}
\usepackage{pstool}
\usepackage{caption}
\usepackage{graphicx}%
\newcommand{\hook}{\text{\large{$\lrcorner$}}}
\usepackage{color}
\usepackage{hyperref}
\newcommand{\qq}[1]{``#1''} 

\newcommand{\di}{\mathrm{d}}
\usepackage{tensor}\newcommand{\ou}[3]{{#1}{}^{#2}{}_{#3}}
\newcommand{\uo}[3]{{#1}{}_{#2}{}^{#3}}

\newcommand{\oepsilon}{\boldsymbol{\hat{\epsilon}}}

\newcommand{\I}{\mathrm{i}} 
\newcommand{\E}{\mathrm{e}} 
\newcommand{\CC}{\mathrm{cc.}} 
\newcommand{\HC}{\mathrm{hc.}} 
\newcommand{\C}{\mathbb{C}}
\newcommand{\N}{\mathbb{N}}
\newcommand{\R}{\mathbb{R}}
\newcommand{\Z}{\mathbb{Z}}

\newenvironment{subalign}{\subequations\align}{\endalign\endsubequations}
\newcommand{\eref}[1]{(\ref{#1})}

\renewcommand{\bold}[1]{\boldsymbol{#1}}

\DeclareMathAlphabet{\sfit}{OT1}{fos}{sb}{it}
\DeclareMathAlphabet{\mathsf}{OT1}{fos}{sb}{n}

\begin{document}

\begin{abstract}
In this article, we study the quantum theory of gravitational boundary modes on a null surface. These boundary modes are given by a spinor and a spinor-valued two-form, which enter the gravitational boundary term for self-dual gravity. Using a Fock representation, we quantise the boundary fields, and show that the area of a two-dimensional cross section turns into the difference of two number operators. The spectrum is discrete, and it agrees with the one known from loop quantum gravity with the correct dependence on the Barbero\,--\,Immirzi parameter. No discrete structures (such as spin network functions, or triangulations of space) are ever required\,---\,the entire derivation happens at the level of the continuum theory. In addition, the area spectrum is manifestly Lorentz invariant.

\end{abstract}
\title{Fock representation of gravitational boundary modes and the discreteness of the area spectrum}
\author{Wolfgang Wieland}
\address{Perimeter Institute for Theoretical Physics\\31 Caroline Street North\\ Waterloo, ON N2L\,2Y5, Canada\\{\vspace{0.5em}\normalfont May 2017}
}
\date{May 2017}
\maketitle
{\tableofcontents}
\begin{center}{\noindent\rule{\linewidth}{0.4pt}}\end{center}
\section{Introduction}

In loop gravity, the quantum states of the gravitational field are built from superpositions of spin network functions, which consist of gravitational Wilson lines for an $SU(2)$ (respectively $SL(2,\C)$) spin connection $\ou{A}{A}{Ba}$. Wherever the Wilson lines meet, all free indices of the parallel transport $\ou{[\mathrm{Pexp}(-\int_\gamma A)]}{A}{B}$ must be saturated and contracted with an invariant tensor (an intertwiner). Otherwise gauge invariance is violated.

\begin{figure}[h]
\begin{center}
\psfrag{z1}{$
\pi^1_A$}
\psfrag{z2}{$\pi^2_A$}
\psfrag{z3}{$\pi^3_A$}
\psfrag{C}{$\mathcal{C}$}
\includegraphics[width=0.35\textwidth]{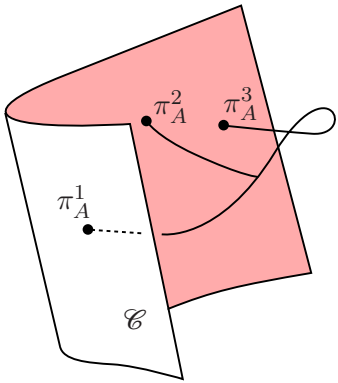}
\end{center}
\caption{In loop quantum gravity the quantum states of the gravitational field are built from gravitational Wilson lines (lying in a three-dimensional spatial hypersurface). These Wilson lines can hit a two-dimensional boundary $\mathcal{C}$, where they create a surface charge, namely a spinor-valued surface density $\pi_A$.
}\label{fig1}
\end{figure}

In the presence of inner boundaries the situation is different.\footnote{The coupling of spin networks to boundaries was first studied in the context of null surfaces that satisfy the isolated horizon boundary conditions \cite{Ashtekar:2000eq,PhysRevD.82.044050,Sahlmann:2011aa}, in our case no such restrictions are required \cite{wieland:nulldefects, Wieland:2017zkf}.} The Wilson lines can now have open ends at the boundary, where they create a surface charge, namely a spinor-valued surface operator $\hat{\pi}_A$. Gauge invariance is restored when both the connection and the boundary spinors transform accordingly. 
Suppose now that there are $N$ such punctures that carry $N$ spinors $\hat{\pi}_A^1,\dots,\hat{\pi}_A^N$ (see \hyperref[fig1]{figure 1}), such that we can introduce the following spinor-valued surface density
\begin{equation}
\hat{\pi}_A(z)=\sum_{i=1}^N\hat{\pi}_A^i{\delta}^{(2)}(z_i,z),\label{densintro}
\end{equation}
where $\delta^{(2)}(\cdot,\cdot)$ is the two-dimensional Dirac distribution at the boundary. In the $N\rightarrow\infty$ continuum and $\hbar\rightarrow 0$ semi-classical limit this surface density will define a classical field $\pi_A(z)$. What is the geometric significance of this surface density in general relativity?

The answer becomes most obvious when considering self-dual (complex) gravity \cite{selfdualtwo}. The action in the bulk is given by the $BF$ topological action plus a constraint, namely
\begin{equation}
S_{\mathcal{M}}[\Sigma,A,\Psi]=\frac{\I}{8\pi G}\int_{\mathcal{M}}\Sigma_{AB}\wedge F^{AB}[A]-\tfrac{1}{2}\Psi^{ABCD}\Sigma_{AB}\wedge\Sigma_{CD},\label{selfdualactn}
\end{equation}
where $\Sigma_{AB}$ is the self-dual Pleba\'nski two-form, $\ou{F}{A}{B}$ is the curvature of the self-dual connection and $\Psi^{ABCD}=\Psi^{(ABCD)}$ is a spin $(2,0)$ Lagrange multiplier (the Weyl spinor) imposing $\Sigma_{(AB}\wedge\Sigma_{CD)}=0$ (the simplicity constraint). If we want to consider a manifold with boundaries (at, say, large but finite distance) boundary terms have to be added, otherwise the variational problem is ill-posed. The defining feature of the self-dual action \eref{selfdualactn} is that all fields carry only unprimed (left-handed) indices $A,B,C,\dots$. Is there then an $SL(2,\C)$ gauge invariant boundary term that has this feature as well (i.e.\ contains only left-handed fields)? In the case of null boundaries, such a boundary term exists \cite{wieland:nulldefects,Wieland:2017zkf}, and its existence relies on the following observation: The pull-back  $\Sigma_{AB\underleftarrow{ab}}$ of the self-dual two-form to a null boundary can be written always as a symmetrised tensor product of a spinor $\ell_A$ and a spinor-valued two-form $\bold{\eta}_{Aab}$, which are both intrinsic to the boundary. An $SL(2,\C)$ gauge invariant boundary term can be then introduced quite immediately, and it is simply given by the three-dimensional boundary integral
\begin{equation}
S_{\partial\mathcal{M}}[\bold{\eta},\ell|A]=\frac{\I}{8\pi G}\int_{\partial\mathcal{M}}\bold{\eta}_A\wedge D\ell^A,\label{bndryintro}
\end{equation}
where $D\ell^A=\di\ell^A+\ou{A}{A}{B}\ell^B$ is the exterior covariant derivative of $\ell^A$. The origin of this boundary term is further explained in \hyperref[sec2]{section 2}, see also \cite{wieland:nulldefects,Wieland:2017zkf} for further references. 

Notice then that the boundary term \eref{bndryintro} is essentially an integral over a symplectic potential. Performing a $2+1$ decomposition along the null boundary, we can identify canonical variables at the boundary. The configuration variable is given by the spinor $\ell^A$ (a null flag), its canonically conjugate momentum is a spinor-valued surface density
\begin{equation}
\pi_A=\frac{\I}{16\pi G}\oepsilon^{ab}\bold{\eta}_{Aab},
\end{equation}
where $\oepsilon^{ab}$ is the Levi-Civita density (a tensor-valued density) on a cross-section of the boundary. 
\vspace{1em}

The purpose of this paper is to give a more thorough analysis of these boundary variables, in both classical and quantum gravity. First of all (\hyperref[sec2]{section} \hyperref[sec2]{2}), we explain the geometric origin of the boundary fields $\ell^A$ and $\bold{\eta}_{Aab}$ for a four-dimensional causal diamond, whose boundary is null (see \hyperref[fig2]{figure 2} for an illustration). Next, we introduce the appropriate boundary and corner terms. We then show that the boundary action \eref{bndryintro} contributes a corner term to the pre-symplectic potential on a space-like three-surface, which intersects the boundary transversally. \hyperref[sec3]{Section 3} deals with the quantum theory. Starting from the boundary fields, we construct four pairs of harmonic oscillators and define the corresponding Fock vacuum. 
Upon introducing the Barbero\,--\,Immirzi parameter, we can then write the \emph{oriented} area of the two-dimensional corner as the difference of two number operators. The spectrum is discrete and matches (up to ordering ambiguities) the loop quantum gravity area spectrum \cite{AshtekarLewandowskiArea,Rovelliarea}. The result is obtained without ever introducing discrete structures, such as spin network functions or triangulations of space. In addition, the derivation is manifestly Lorentz invariant. Finally (\hyperref[sec4]{section 4}), we explain the compatibility of the result with loop gravity in the spin network representation. 

\vspace{1em} 
The paper is part of a wider effort \cite{Wieland:2016dbc,Wieland:2017zkf,wieland:nulldefects} to understand null surfaces, causal structures and internal boundaries in non-perturbative and canonical quantum gravity in terms of the spinorial representation of loop quantum gravity \cite{twist,komplexspinors,Bianchi:2016hmk}. A similar formalism using metric variables (rather than spinors) is being developed by Freidel and collaborators \cite{Donnelly:2016auv,Freidel:2015gpa}, see also \cite{Geiller:2017xad} for gravity in three dimensions. In addition, our results are probably relevant for the so-called $BF$ representation  \cite{Dittrich:2014wpa,Bahr:2015bra} as well.\footnote{Bahr, Dittrich and Geiller \cite{Dittrich:2014wpa,Bahr:2015bra} have proposed recently a radical reformulation of loop quantum gravity in the continuum, which is built over a distributional vacuum peaked at flat or constantly curved three-geometries. In this new representation, one finds a more complicated area spectrum \cite{Bahr:2015bra}. In our continuous Fock representation, the original loop gravity area spectrum is recovered (up to quantisation ambiguities). The two representations are therefore likely  unitarily inequivalent, such that normalised states in one representation may only reappear as distributions in the other. }

\section{Boundary and corner terms for self-dual variables}\label{sec2}
Consider then general relativity in the self-dual formulation \cite{selfdualtwo}. The configuration variables in the bulk are the self-dual Pleba\'nski area two-form $\Sigma_{AB}=\Sigma_{(AB)}$ and the $\mathfrak{sl}(2,\C)$ connection $\ou{A}{A}{B}$ with curvature $\ou{F}{A}{B}=\di\ou{A}{A}{B}+\ou{A}{A}{C}\wedge\ou{A}{C}{B}$. Physical motions are given by those field configurations that extremise the topological $BF$ action
\begin{equation}
S_{\mathcal{M}}[\Sigma, A]=\left[\frac{\I}{8\pi G}\frac{\beta+\I}{\beta}\int_{\mathcal{M}}\Sigma_{AB}\wedge F^{AB}\right]+\CC\label{BFactn}
\end{equation}
in the class of all fields that satisfy the simplicity constraints
\begin{subequations}\begin{align}
&\Sigma_{(AB}\wedge\Sigma_{CD)}\stackrel{!}{=}0,\\
&\Sigma_{AB}\wedge\bar\Sigma_{A'B'}\stackrel{!}{=}0,\\
&\Sigma_{AB}\wedge\Sigma^{AB}+\bar{\Sigma}_{A'B'}\wedge\bar{\Sigma}^{A'B'}\stackrel{!}{=}0.
\end{align}\label{simplcons}\end{subequations}
The simplicity constraints guarantee that the area two-form $\Sigma_{AB}$ is compatible with the existence of a Lorentzian metric $g_{ab}=\epsilon_{AB}\bar\epsilon_{A'B'}\ou{e}{AA'}{a}\ou{e}{BB'}{b}$ for 
a tetrad $\ou{e}{AA'}{a}=-\ou{\bar{e}}{A'A}{a}$, such that either\footnote{We will later restrict ourselves to only one of these four solution sectors, namely the first $\Sigma_{AB}=-\frac{1}{2}e_{AC'}\wedge\uo{e}{B}{C'}$, which corresponds to $\Sigma_{AA'BB'}= -\bar\epsilon_{A'B'}\Sigma_{AB}-\epsilon_{AB}\bar\Sigma_{A'B'}= e_{AA'}\wedge e_{BB'}$ and a signature $($$-$$+$$+$$+)$ metric $g_{ab}$. \label{foot2} }
\begin{equation}\begin{split}
\Sigma_{AB}&=\mp\frac{1}{2}e_{AC'}\wedge \uo{e}{B}{C'},\quad\text{or}\\
\Sigma_{AB}&=\mp\frac{\I}{2}e_{AC'}\wedge \uo{e}{B}{C'}.\end{split}\label{selfsolv}
\end{equation}
 The equations of motion for any one of these solutions are then the torsionless condition,
\begin{equation}
\nabla\ou{\Sigma}{A}{B}=0\Leftrightarrow\nabla_{[a}\ou{\Sigma}{AB}{ab]}=0\label{torsnless}
\end{equation}
and the Einstein equations, which demand that the curvature be Ricci flat
\begin{equation}
\ou{F}{A}{B}=\ou{\Psi}{A}{BCD}\Sigma^{CD},\label{Eeq}
\end{equation}
where $\nabla=\di+[A,\cdot]$ is the exterior covariant derivative and $\Psi_{ABCD}=\Psi_{(ABCD)}$ is the spin $(2,0)$ Weyl spinor. 

The action \eref{BFactn} contains two coupling constants, namely Newton's constant $G$, which is a mere conversion factor between units of action and units of area (for $\hbar=c=1$), and the Barbero\,--\,Immirzi parameter $\beta$, which is a pure number ($\beta>0$). The addition of the Barbero\,--\,Immrizi parameter is actually necessary: Had we not introduced $\beta$, and worked with the action \eref{BFactn} for $\beta\rightarrow\infty$ (or $\beta\rightarrow 0$) instead, the equations of motion would be less restrictive: Any Lorentzian manifold would be a stationary point of the action.\footnote{This can be seen as follows: If $\Sigma_{AB}$ is a solution of the simplicity constraints \eref{simplcons}, we can build a new such solution simply by replacing $\Sigma_{AB}$ by $
\I\times\Sigma_{AB}$. At the level of the self-dual variables, multiplication by the imaginary unit amounts to take the Hodge dual in the internal indices $\alpha,\beta,\gamma,\dots$. If the simplicity constraints are satisfied, there exists then a tetrad $e^\alpha$ such that either $\Sigma_{AB}$ (case i) or $\I\times\Sigma_{AB}$ (case ii) is given by the self-dual part of $\pm e_\alpha\wedge e_\beta$, see \cite{selfdualtwo}. If we now insert any such $\Sigma_{AB}$ for e.g.\ $\beta\rightarrow\infty$ back into the action, we are left with the Einstein\,--\,Hilbert action in the first case, but in the second case we only get a topological term, namely $1/16\pi G\int_{\mathcal{M}}e_\alpha\wedge e_\beta\wedge F^{\alpha\beta}$ (equally for $\beta\rightarrow0$ and $\beta G=\mathrm{const.}$). The resulting equations of motion would be the torsionless condition $\nabla e^\alpha=0$ alone, and there would be no Einstein equations, since the variation of $e^\alpha$ would only yield the Bianchi identity $\ou{F}{\alpha}{\beta}\wedge e^\beta=0$, which is already a consequence of the vanishing of torsion $T^\alpha=\nabla e^{\alpha}=0$.}

\begin{figure}[h]
\begin{center}
\psfrag{A}{$\varSigma_-$}
\psfrag{B}{$\varSigma_+$}
\psfrag{C}{$\mathcal{N}_+$}
\psfrag{D}{$\mathcal{N}_-$}
\psfrag{E}{$\mathcal{\mathcal{C}}_o$}
\psfrag{F}{$\mathcal{C}_-$}
\psfrag{G}{$\mathcal{C}_+$}
\includegraphics[width=0.46\textwidth]{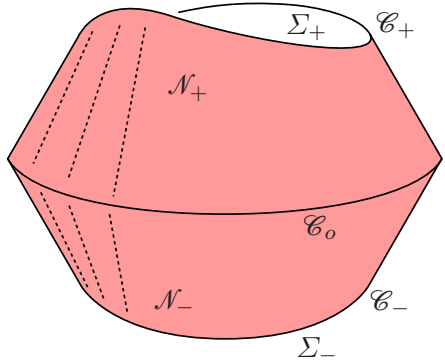}
\end{center}
\caption{We are considering the gravitational field in a four-dimensional causal region $\mathcal{M}$, whose boundary has four components, namely the three-dimensional null surfaces $\mathcal{N}_+$ and $\mathcal{N}_-$, which have the topology of a cylinder $[0,1]\times S^2$, and the spacelike disks $\varSigma_-$ and $\varSigma_+$ at the top and bottom. The boundary has three corners, which appear as the boundary of the boundary, namely $\partial\mathcal{N}_+=\mathcal{C}_+\cup\mathcal{C}_o^{-1}$ and $\partial\mathcal{N}_-=\mathcal{C}_o\cup\mathcal{C}_-^{-1}$. All these manifolds carry an orientation, which is induced from the bulk: $\partial\mathcal{M}={\varSigma}_{-}^{-1}\cup\mathcal{N}_-\cup\mathcal{N}_+\cup\varSigma_+$.}\label{fig2}
\end{figure}

\vspace{1em}
We are then considering the gravitational field in a compact four-dimen\-sional causal diamond $\mathcal{M}$ as drawn in \hyperref[fig2]{figure 2}. The boundary $\partial\mathcal{M}$ consists of four components: Three-dimensional null surfaces $\mathcal{N}_+$ and $\mathcal{N}_-$, and spacelike hypersurfaces $\varSigma_+$ and $\varSigma_-$ at the top and bottom cutting off the diamond 
 before the null surfaces recollapse into a point or caustic. 

Next, we have to add boundary and corner terms such that the variational problem is well-posed. For self-dual variables on a null surfaces, these boundary terms have been studied in the earlier papers \cite{Wieland:2017zkf,wieland:nulldefects} of the series. The boundary action is built from certain boundary fields, which are intrinsic to a null surface: On a null surface $\mathcal{N}$ there always exists, in fact, a spinor-valued two-form $\ou{\bold{\eta}}{A}{ab}\in\Omega^2(\mathcal{N}:\C^2)$ and a two-component Weyl spinor (a spinor-valued $0$-form) $\ell^A\in\Omega^0(\mathcal{N}:\C^2)$ such that the pull-back $\varphi^\ast_{\mathcal{N}}:T^\ast\mathcal{M}\rightarrow T^\ast\mathcal{N}$ of the self-dual two-form $\ou{\Sigma}{AB}{ab}$ to the null boundary turns into the symmetrised spin $(1,0)$ tensor product
\begin{equation}
\left[\varphi^\ast_{\mathcal{N}}\Sigma^{AB}\right]_{ab}\equiv\ou{\Sigma}{AB}{\underleftarrow{ab}}=\ou{\bold{\eta}}{(A}{ab}\ell^{B)}.\label{glucond}
\end{equation}
It can be then shown (see again \cite{Wieland:2017zkf,wieland:nulldefects} for the details) that the boundary spinors $(\bold{\eta}_{Aab},\bar{\ell}^{A'})$ determine the entire intrinsic geometry\footnote{The intrinsic geometry is determined completely by a degenerate signature $(0$$+$$+)$ metric $q_{ab}$, whose degenerate direction determines the direction $[\ell^a]\ni \ell^a:q_{ab}\ell^b=0$ of the null generators.} of the null surface. For instance, there is the spin $(\tfrac{1}{2},\tfrac{1}{2})$ vector component
\begin{equation}
\ell^\alpha\equiv\I\ell^A\bar{\ell}^{A'},
\end{equation}
and it determines the internal null surface generators $\ell^\alpha=\ou{e}{\alpha}{a}\ell^a$. The spin zero singlet
\begin{equation}
\bold{\varepsilon}_{ab}=-\I\bold{\eta}_{Aab}\ell^A,\label{areatwoform}
\end{equation}
on the other hand, defines the area two-form, which is intrinsic to the boundary. In fact, the two-dimensional and oriented area of any two-dimensional cross-section of the boundary is given by the integral
\begin{equation}
\mathrm{Ar}[\mathcal{C}]=-\I\int_{\mathcal{C}}\bold{\eta}_{A}\ell^A.\label{Ardef}
\end{equation}
For the area to be real constraints must be satisfied, namely the \emph{reality conditions}\footnote{It is here that we restrict ourselves to only the first solution sector \eref{selfsolv} of the simplicity constraints \eref{simplcons}. See also \hyperref[foot2]{footnote 2}.}
\begin{equation}
\bold{\eta}_{Aab}\ell^A+\CC=0.\label{realcond}
\end{equation}
The action for the entire region consists then of the action \eref{BFactn} in the bulk plus a boundary and corner term, namely 
\begin{align}\nonumber
S[A,\Sigma&,\bold{\eta},\ell,\bold{\alpha}]=\\\nonumber
=&\frac{\I}{8\pi G}\frac{\beta+\I}{\beta}\bigg[\int_{\mathcal{M}}\Sigma_{AB}\wedge F^{AB}
+\int_{\mathcal{N}_+}\bold{\eta}^+_A\wedge\big((D-\omega)\ell^A_+-\psi^A_+\big)+\\
&+\int_{\mathcal{N}_-}\bold{\eta}^-_A\wedge\big((D+\omega)\ell^A_-+\psi^A_-\big)+\int_{\mathcal{C}_o}\bold{\alpha}(\ell^-_A\ell^A_+-1)\bigg]+\CC
\end{align}
where $D_a=\nabla_{\underaccent{\leftarrow}{a}}$ is the pull-back of the exterior covariant derivative from the bulk to the boundary. The action is to be extremised in the class of all fields that satisfy the simplicity constraints \eref{simplcons} for given boundary conditions
\begin{subalign}
&\text{on $\mathcal{N}_\pm$:} 
\quad\delta\omega_a=0,\quad\delta[\psi^A_\pm]_a=0,\\
&\text{on $\mathcal{\varSigma}_\pm$:}\quad\delta\ou{A}{A}{B\underaccent{\leftarrow}{a}}=0,\\
&\text{on $\mathcal{\mathcal{C}}_\pm$:}\quad\delta\ell^A_\pm=0.
\end{subalign}
 The resulting equations of motion are the Einstein equations \eref{Eeq} and the torsionless condition \eref{torsnless} in the bulk. At the null boundary, additional boundary equations of motion appear: The variation of the boundary spinors determines the exterior covariant derivatives $D_a\ell^A$ and $D_{[a}\ou{\bold{\eta}}{A}{bc]}$ in terms of the external potentials\footnote{The paper \cite{Wieland:2017zkf} explains the geometric significance of $\omega_a$ and $\ou{\psi}{A}{a}$ as a measure for the extrinsic curvature of the null boundary.} $\omega_a$ and $\ou{\psi}{A}{a}$, which are held fixed in the variational problem. In addition, the variation of the action also determines the glueing conditions, which link the boundaries and corners with the variables in the bulk, namely
\begin{subalign}
\big[\varphi^\ast_{\mathcal{N}_\pm}\Sigma_{AB}\big]_{ab}&=\ell^\pm_{(A}\bold{\eta}^\pm_{B)ab},\label{glu1}\\
\big[\varphi^\ast_{\mathcal{C}_o}\bold{\eta}^\pm_{A}\big]_{ab}&=\ell_A^\pm\bold{\alpha}_{ab}\label{glu2},
\end{subalign}
where e.g.\ $\varphi^\ast_{\mathcal{C}}$ is the pull-back $\varphi^\ast_{\mathcal{C}}:T^\ast\mathcal{N}\rightarrow T^\ast\mathcal{C}_o$. Equation \eref{glu1} is obtained from the variation of the connection along the null surface, whereas \eref{glu2} follows from the variation of $\ell^A_\pm$ at the intersection $\mathcal{C}_o=\mathcal{N}_+\cap\mathcal{N}_-$. Finally, there is also the variation with respect to the two-form $\bold{\alpha}$ at the corner $\mathcal{C}_o=\mathcal{N}_+\cap\mathcal{N}_-$, and it simply fixes the normalisation  $\ell_A^-\ell^A_+=1$ of the spin dyad $(\ell^A_-,\ell^A_+)$ at the corner.

 The variation of the action determines both the equations of motion and the covariant symplectic potential (at the pre-symplectic or kinematical level), namely
\begin{equation}
\delta S=\mathrm{EOM}\cdot\delta+\Theta_{\partial\mathcal{M}}(\delta).
\end{equation}
For each one of the boundary components, there is then a term in the pre-symplectic potential, namely
\begin{subequations}
\begin{align}
\Theta_{\varSigma_\pm}&=\frac{\I}{8\pi G}\frac{\beta+\I}{\beta}\left[\int_{\varSigma_{\pm}}\Sigma_{AB}\wedge\bbvar{d}{A}^{AB}+\int_{\mathcal{C}_\pm}\bold{\eta}_A^\pm\bbvar{d}\ell^A_\pm\right]+\CC,\label{symppot1}\\
\Theta_{\mathcal{N}_\pm}&=\mp\frac{\I}{8\pi G}\frac{\beta+\I}{\beta}\int_{\mathcal{N}_\pm}\Big[\bold{\eta}_A^\pm\ell^A_\pm\wedge\bbvar{d}\omega+\bold{\eta}_A^\pm\wedge\bbvar{d}\psi^A_\pm\Big]+\CC
\end{align}
\end{subequations}
In the following, we will  restrict ourselves to only one such component, namely ${\varSigma_+}\equiv\varSigma$. The symplectic potential $\Theta_\varSigma$ consist of a three-dimensional integral over the interior, and an additional two-dimensional integral over the corner $\mathcal{C}\equiv\mathcal{C}_+$. The goal of the remaining part of the paper is to study the quantisation of the phase space at this two-dimensional corner alone. The canonical analysis of the entire phase space including the new boundary variables $\bold{\eta}_{Aab}^\pm$ and $\ell^A_\pm$ will be left to a forthcoming publication in this series. The approach so far is therefore incomplete: we will quantise the symplectic structure at the corner, but we will leave the degrees of freedom in the bulk classical.

\section{Landau quantisation of area}\label{sec3}
\subsection{Landau operators}
In this section, we will develop our main result, namely a new representation of quantum geometry  that reproduces the discrete loop quantum gravity area spectrum in the continuum, without ever relying on a discretisation of space,  lattice variables or a gauge fixing to a compact gauge group. In addition, our construction is manifestly Lorentz invariant. 

Our starting point is the classical phase space at the corner. The Poisson brackets for the boundary variables are determined by the corner term
\begin{equation}
\frac{\I}{8\pi G}\frac{\beta+\I}{\beta}\int_{\mathcal{C}}\big(\bold{\eta}_A\bbvar{d}\ell^A-\CC\big)
\end{equation}
appearing in the symplectic potential \eref{symppot1}. The spinor $\ell^A$ plays the role of the configuration variable. Its conjugate momentum is given by the spinor-valued surface density 
\begin{equation}
\pi_{A}:=\frac{\I}{16\pi G}\frac{\beta+\I}{\beta}\oepsilon^{ab}\bold{\eta}_{Aab},
\end{equation}
where $\beta>0$ denotes the Barbero\,--\,Immirzi parameter and $\oepsilon^{ab}$ is the two-dimensional and metric-independent Levi-Civita density at the corner.\footnote{If $\{\vartheta^1,\vartheta^2\}$ are coordinates on $\mathcal{C}$, this density is defined by $\oepsilon^{ab}=\di\vartheta^i\wedge\di\vartheta^j\frac{\partial^a}{\partial\vartheta^i}\frac{\partial^b}{\partial\vartheta^j}.$} The fundamental Poisson brackets are given by
\begin{subequations}\begin{align}
\big\{\pi_{A}(z),\ell^B(z')\big\}_{\mathcal{C}}&=\delta^B_A\delta^{(2)}(z,z'),
\\
\big\{\bar{\pi}_{A'}(z),\bar{\ell}^{B'}(z')\big\}_{\mathcal{C}}&=\delta^{B'}_{A'}\delta^{(2)}(z,z'),
\end{align}\label{spinPoissn}\end{subequations}
where $\delta^{2}(\cdot,\cdot)$ is the two-dimensional Dirac distribution at the corner. All other Poisson brackets among the canonical variables vanish identically.
\vspace{1em}

The spinors $\ell^A$ and $\pi_{A}$ are not arbitrary. The reality conditions \eref{realcond} constrain the spin $(0,0)$ singlet $\pi_A\ell^A$ to satisfy
\begin{equation}
C=\frac{\I}{\beta+\I}\pi_{A}\ell^A+\CC=0.\label{realcond2}
\end{equation}
The reality conditions are necessary for the spinors to be compatible with a real and Lorentzian metric in a neighbourhood of the corner. On the  $C=0$ constraint hypersurface in phase space, we can then find the following identities for the area in terms of the canonical variables, namely
\begin{equation}
\mathrm{Ar}[\mathcal{C}]=-\I\int_{\mathcal{C}}\bold{\eta}_A\ell^A\approx\frac{1}{2\I}\int_{\mathcal{C}}\big(\bold{\eta}_A\ell^A-\CC\big)\approx4\pi\I\,\beta G\int_{\mathcal{C}}(\pi_A\ell^A-\CC).
\end{equation}
where \qq{$\approx$} means equality up to terms that vanish for $C=0$. The right hand side is clearly real and well-defined on the entire phase space. Defining
\begin{equation}
\mathrm{Ar}[\mathcal{C}]:=4\pi\I\,\beta G\int_{\mathcal{C}}(\pi_A\ell^A-\CC),\label{areaop}
\end{equation}
we can extend, therefore, the definition of the area away from the $C=0$ constraint hypersurface, thus turning the area into a partial observable \cite{partobs} on the entire kinematical phase space over the corner. 

To quantise the theory, we construct harmonic oscillators from $\pi_A$ and $\ell^A$. We will then define the canonical Fock space for these oscillators and compute the spectrum of the area operator \eref{areaop} at the quantum level. To define harmonic oscillators, we need, however, additional geometrical background structures at the corner. An example illustrates the situation: Consider a particle in a complex plane, $z=x+\I y$ is the position, $p=1/2(p_x-\I p_y)$ denotes the conjugate momentum. The Poisson brackets are  $\{p_z,z\}=\{\bar{p}_z,\bar{z}\}=1$. The Landau operators $a:=\sqrt{\Omega/2}\,(\bar{z}+\I\Omega^{-1} p_z)$ and $b:=\sqrt{\Omega/2}\,(z+\I\Omega^{-1}\bar{p}_{z})$ are then built by taking the sum of the configuration variable (which is $\ell^A$ in our case) and the complex conjugate momentum variable (which is $\bar{\pi}_{A'}$). Notice now that the complex coordinates $(a, b)$ depend on an additional length scale, namely $\Omega$, which is required because $p_z$ and $z$ have opposite dimensions of length. The same happens for $(\pi_A,\bar\ell^{A'})$. The momentum variable is a density weight one spinor, to sum $\bar{\pi}_{A'}$ with $\ell^A$ we need to first divide by an appropriate surface density $d^2\Omega$, and then map primed into unprimed indices before taking the sum of $\ell^A$ and $\bar{\pi}_{A'}$. We therefore need \emph{two} additional structures: a two-dimensional fiducial volume element $d^2\Omega$ and a complex structure (essentially a Hermitian metric) mapping primed into unprimed indices $A'\rightarrow A$.  

Accordingly, we choose a fiducial and non-degenerate area two-form ${}^\circ\!\bold{\varepsilon}_{ab}\in\Omega^2(\mathcal{C}:\R_>)$, such that the surface density
\begin{equation}
d^2\Omega=\frac{1}{2}\oepsilon^{ab}{}^\circ\!\bold{\varepsilon}_{ab}=\Omega^2(\vartheta,\varphi)\,\di\hspace{0.04em}\mathrm{cos}\vartheta\wedge\di\varphi,
\end{equation}
is positive ($\vartheta$ and $\varphi$ are spherical coordinates with respect to some fiducial round background metric $\delta=\di\vartheta^2+\mathrm{sin}^2\vartheta\,\di\varphi^2$ at the corner). We will then also need the inverse ${}^\circ\!\bold{\varepsilon}^{ab}$ of $d^2\Omega$ (a section of the anti-symmetric tensor bundle  $T\mathcal{N}\wedge T\mathcal{N}$ over the corner), which is defined implicitly by 
\begin{equation}
{}^\circ\!\bold{\varepsilon}^{ac}\,{}^\circ\!\bold{\varepsilon}_{bc}=[\mathrm{id}_{\mathcal{N}}]^a{}_b,
\end{equation}
where $\mathrm{id}_{\mathcal{N}}:T\mathcal{N}\rightarrow T\mathcal{N}$ is the identity.

Next, we choose\footnote{The most natural choice is given by the surface normal $n^a$ of $\varSigma$ itself, such that $n^\alpha=\ou{e}{\alpha}{a}n^a$. Notice that this turns $n^\alpha$ into a field-dependent and internal four-vector, which depends (as a functional) on the pull-back of the tetrad to $\varSigma$.} an internal and future oriented normal $n^\alpha:\eta_{\alpha\beta}n^\alpha n^\beta=-1$ such that we have a Hermitian metric in the spin bundle over the corner, namely
\begin{equation}
\delta_{AA'}=\sigma_{AA'\alpha}n^\alpha,
\end{equation}
where $\sigma_{AA'\alpha}$ are the internal and four-dimensional soldering forms.\footnote{A  matrix representation is given by the four-dimensional Pauli matrices $\ou{\sigma}{AA'}{\alpha}=(\bbvar{1},\sigma_1,\sigma_2,\sigma_3)$. The relation between the tetrad is given by $\ou{e}{AA'}{a}=\frac{\I}{\sqrt{2}}\ou{\sigma}{AA'}{\alpha}\ou{e}{\alpha}{a}.$} We then have a norm and can set
\begin{subalign}
\|\ell\|^2=\delta_{AA'}\ell^A\bar{\ell}^{A'},\qquad\|\pi\|^2=\delta^{AA'}\pi_A\bar{\pi}_{A'}.
\end{subalign}
Notice that $\|\ell\|^2$ is a scalar, whereas $\|\pi\|^2$ is a surface density of weight four.

Having introduced both a fiducial volume element (the surface density $d^2\Omega$) and a complex structure (the Hermitian metric $\delta_{AA'}=\sigma_{AA'\alpha}n^\alpha$), we can introduce now the Landau operators
\begin{subequations}\begin{align}
a^A[d^2\Omega,n^\alpha]\equiv a^A&=\frac{d^{\frac{2}{2}}\Omega}{\sqrt{2}}\left(\delta^{AA'}\bar\ell_{A'}-\frac{\I}{2}{}^\circ\!\bold{\varepsilon}^{ab}\ou{\pi}{A}{ab}\right),\label{harmop1}\\
b^A[d^2\Omega,n^\alpha]\equiv b^A&=\frac{d^{\frac{2}{2}}\Omega}{\sqrt{2}}\left(\ell^A+\frac{\I}{2}\delta^{AA'}{}^\circ\!\bold{\varepsilon}^{ab}\bar{\pi}_{A'ab}\right),\label{harmop2}
\end{align}\label{Landauop}\end{subequations}
which are spinor-valued half-densities that depend parametrically on the fiducial background structures $n^\alpha$ (which is an internal, future oriented normalised four-vector) and $d^{\frac{2}{2}}\Omega$ (which is the half-density  $\sqrt{d^2\Omega}$). The fundamental Poisson commutation relations \eref{spinPoissn} translate now into commutation relations for two pairs of harmonic oscillators over the sphere, namely
\begin{subalign}
\big\{a^A(z),a^\ast_B(z')\big\}_{\mathcal{C}}=\I\,\delta^A_B\,\delta^{(2)}(z,z'),\\
\big\{b^A(z),b^\ast_B(z')\big\}_{\mathcal{C}}=\I\,\delta^A_B\,\delta^{(2)}(z,z'),
\end{subalign}
where we introduced the conjugate spinors
\begin{equation}
a^\ast_A=\delta_{AA'}\bar{a}^{A'},\qquad b^\ast_A=\delta_{AA'}\bar{b}^{A'}.\label{Fvac}
\end{equation}

In quantum theory, the Fock vacuum $|0,\{d^2\Omega,n_\alpha\}\rangle$ is then given as the state in the kernel of the annihilation operators,
\begin{equation}
\forall z\in\mathcal{C}:\hat{a}^A(z)\big|0,\{d^2\Omega,n_\alpha\}\big\rangle=\hat{b}^A(z)\big|0,\{d^2\Omega,n_\alpha\}\big\rangle=0.\label{Fockvac}
\end{equation}
Next, we introduce the canonical number operators for the two oscillators over any point in $\mathcal{C}$, namely
\begin{subequations}\begin{align}
N_a=a^\ast_A a^A&=\frac{1}{2}\Big[d^2\Omega\|\ell\|^2+(d^2\Omega)^{-1}\|\pi\|^2+\I(\ell^A\pi_A-\bar{\pi}_{A'}\bar{\ell}^{A'})\Big],\label{numbop1}\\
N_b=\,b^\ast_A b^A&=\frac{1}{2}\Big[d^2\Omega\|\ell\|^2+(d^2\Omega)^{-1}\|\pi\|^2-\I(\pi_A\ell^A-\bar{\ell}^{A'}\bar{\pi}_{A'})\Big].\label{numbop2}
\end{align}\label{numbop}\end{subequations}
We can then also introduce the squeeze operators
\begin{subequations}\begin{align}
a_Ab^A&=-\frac{1}{2}\Big[d^2\Omega\|\ell\|^2-(d^2\Omega)^{-1}\|\pi\|^2+\I(\bar\ell^{A'}\bar{\pi}_{A'}+\pi_A\ell^A)\Big],\label{squeezop1}\\
(a_Ab^A)^\ast&
=-\frac{1}{2}\Big[d^2\Omega\|\ell\|^2-(d^2\Omega)^{-1}\|\pi\|^2-\I(\bar{\ell}^{A'}\bar\pi_{A'}+\pi_A\ell^A)\Big].\label{squeezop2}
\end{align}\label{squeezop}\end{subequations}

\subsection{Quantisation of area}
For the purpose of this paper, the two most relevant operators are the area operator \eref{areaop} and the reality conditions \eref{realcond}. Choosing a normal ordering, the area operator is nothing but the difference of the two number operators, namely
\begin{align}\nonumber
\boldsymbol{\colon}\!{\mathrm{Ar}}[\mathcal{C}]\boldsymbol{\colon}&=2\pi\I\,\beta G\int_{\mathcal{C}}(\hat{\pi}_A\hat{\ell}^A+\hat{\ell}^A\hat{\pi}_A-\mathrm{h.c.})=\\
&=4\pi\beta G\int_{\mathcal{C}}\big(\hat{a}^\dagger_A\hat{a}^A-\hat{b}^\dagger_A\hat{b}^A\big).\label{areaopdef}
\end{align}
The spectrum of this operator in the Fock space over the vacuum \eref{Fvac} is discrete. This becomes particularly obvious if we introduce the following basis: Consider spinor spherical harmonics\footnote{The spinor spherical harmonics are defined with respect to some fiducial two-dimensional round metric $\delta=\di\vartheta^2+\mathrm{sin}^2\vartheta\,\di\varphi^2$ at the corner $\mathcal{C}$.} $Y^A_{JMs}(\vartheta,\varphi)$, such that we have a mode expansion
\begin{subalign}
\hat{a}^A(\vartheta,\varphi)=d^{\frac{2}{2}}\Omega\sum_{J=1/2}^\infty\sum_{M=-J}^{J}\sum_{s=\pm}\hat{a}_{JMs}Y^A_{JMs}(\vartheta,\varphi),\\
\hat{b}^A(\vartheta,\varphi)=d^{\frac{2}{2}}\Omega\sum_{J=1/2}^\infty\sum_{M=-J}^{J}\sum_{s=\pm}\hat{b}_{JMs}Y^A_{JMs}(\vartheta,\varphi).
\end{subalign}
The resulting Poisson brackets are given by an infinite tower of harmonic oscillators,
\begin{equation}
\big[\hat{a}_{JMs},\hat{a}^\dagger_{J'M's'}\big]=\big[\hat{b}_{JMs},\hat{b}^\dagger_{J'M's'}\big]=\delta_{JJ'}\delta_{MM'}\delta_{ss'},
\end{equation}
where we chose the canonical normalisation
\begin{equation}
\int_{\mathcal{C}}d^2\Omega\,\delta_{AA'}{\bar{Y}}{}^{A'}_{J'M's'} Y^A_{JMs}=\delta_{JJ'}\delta_{MM'}\delta_{ss'}.
\end{equation}
The area operator is now just the sum of the differences of the two number operators in each mode, namely
\begin{equation}
\boldsymbol{\colon}\!{\mathrm{Ar}}[\mathcal{C}]\boldsymbol{\colon}=4\pi\beta\, G\sum_{J={1}/{2}}^\infty\sum_{M=-J}^J\sum_{s=\pm1}\left(\hat{a}^{{\dagger}}_{JMs}\hat{a}^{\phantom{\dagger}}_{JMS}-\hat{b}^\dagger_{JMs}\hat{b}^{\phantom{\dagger}}_{JMs}\right).
\label{ardef0}\end{equation}
Hence, there is a fundamental discreteness of area in quantum gravity. The possible eigenvalues $\{a_n\}$ of area are given by the multiplies
\begin{equation}
a_n=4\pi\beta\, G\,n={a}_o\frac{n}{2},\quad n\in\Z\label{arspec}
\end{equation}
of the fundamental loop gravity area gap
\begin{equation}
a_o=8\pi\beta\, G=8\pi\beta\,\ell_{\mathrm{P}}^2,\label{areagap}
\end{equation}
where $\beta>0$ is the Barbero\,--\,Immirzi parameter and $\ell_{\mathrm{P}}=\sqrt{\hbar G/c^3}$ is the Planck length. Notice that the area spectrum contains both positive and negative eigenvalues. This is to be expected, since the classical expression \eref{Ardef} measures the oriented area of the cross-section, which may be positive or negative depending on the orientation. The area spectrum is equidistant, and it differs, therefore, from the one that has been found in loop quantum gravity. There is no logical contradiction: In here, we are quantising a different operator, namely the oriented area of a two-dimensional cross section of a null surface, whereas in loop quantum gravity one studies the metrical area
\begin{equation}
\mathrm{Ar}_{g}[\mathcal{C}]=\int_{\mathcal{C}}\di x\,\di y\,\sqrt{\mathrm{det}\begin{pmatrix}g(\partial_x,\partial_x)&g(\partial_x,\partial_y)\\
g(\partial_y,\partial_x)&g(\partial_y,\partial_y)\end{pmatrix}}\label{ardef1}
\end{equation}
instead. In fact, in loop gravity, one finds \cite{AshtekarLewandowskiArea,Rovelliarea} the following main eigenvalues for the metrical area of a surface, namely
\begin{equation}
a_{n_1,n_2,\dots}=8\pi\beta\,G\,\sum_{2j\in\mathbb{N}} n_{2j}\sqrt{j(j+1)},\quad n_i\in\N_0.
\end{equation}
At the classical level, both notions of area agree up to a local sign, whereas in quantum theory, the eigenvalues of the two operators \eref{ardef1} and \eref{ardef0} are different, but they approach each other in the semi-classical limit, namely for $j\rightarrow\infty$, $\hbar j=\mathrm{const.}$ 
\subsection{Imposition of the reality conditions}
Finally, there are also the reality conditions \eref{realcond}, which we now need to impose at the quantum level as well. Separating real and imaginary parts of $\pi_A\ell^A$, we get
\begin{align}\nonumber
C&=\frac{\I}{\beta+\I}\pi_A\ell^A+\CC=\\
&=\frac{1}{\beta^2+1}\left[(\pi_A\ell^A+\bar{\ell}^{A'}\bar{\pi}_{A'})+\I\beta(\pi_A\ell^A-\bar{\ell}^{A'}\bar{\pi}_{A'})\right].
\end{align}
Choosing a normal ordering, and going back to the definition of the number and squeeze operators (\ref{squeezop1}, \ref{squeezop2}), we can quantise this operator rather immediately, namely by saying
\begin{align}
\boldsymbol{\colon}\!C\boldsymbol{\colon}=\frac{1}{\beta^2+1}\left[\I\big(\hat{a}_A\hat{b}^A-(\hat{a}_A\hat{b}^A)^\dagger\big)+\beta\big(\hat{a}^\dagger_A\hat{a}^A-\hat{b}^\dagger_A\hat{b}^A\big)\right].\label{realcondfock}
\end{align}
A short calculation confirms that the area operator \eref{areaopdef} commutes with the quantum reality conditions \eref{realcondfock}, namely that
\begin{equation}
\forall z\in\mathcal{C}:\Big[\boldsymbol{\colon}\!C(z)\boldsymbol{\colon},\boldsymbol{\colon}\!{\mathrm{Ar}}[\mathcal{C}]\boldsymbol{\colon}\hspace{-0.3em}\Big]=0.
\end{equation}
In the same way, one can also show that the area operator and the reality conditions \eref{realcondfock} commute with the generators of local $SL(2,\C)$ gauge transformations, which are given by
\begin{subequations}\begin{align}
\Pi_{AB}(z)&=-\frac{1}{2}\hat{\pi}_{(A}(z)\hat{\ell}_{B)}(z),\\
\bar\Pi_{A'B'}(z)&=-\frac{1}{2}\hat{\pi}^\dagger_{(A'}(z)\hat{\ell}^\dagger_{B')}(z).
\end{align}\label{Lorentzgen}\end{subequations}
Notice that there is no ordering ambiguity in here, because $[\hat{\pi}_A(z),\ell_B(z')]=-\I\epsilon_{AB}\delta^{(2)}(z,z')$ is anti-symmetric in $A$ and $B$, whereas $\Pi_{AB}=\Pi_{BA}$ is symmetric.

In summary, the area operator, the reality conditions \eref{realcondfock} and the generators of local $SL(2,\C)$ gauge transformations can be diagonalised simultaneously. In \eref{arspec} we gave the spectrum of the area operator at the kinematical level (i.e.\ prior to imposing the reality conditions). The area operator commutes with both the reality conditions \eref{realcondfock} and the $SL(2,\C)$ generators, and the spectrum at the level of the physical (or gauge invariant) boundary Hilbert space can therefore only be a subset of the kinematical area spectrum. \vspace{1em}

To impose the reality conditions \eref{realcondfock} at the quantum level, we first introduce the corresponding finite gauge transformations $U[\lambda]:= \exp(-\I\int_{\mathcal{C}}\lambda C)$ for gauge parameters $\lambda:\mathcal{C}\rightarrow\R$. Any such gauge transformation generates a conformal transformation of the fiducial area element $d^2\Omega$ in addition to a local $U(1)$ phase rotation. This can be seen by writing the constraint \eref{realcondfock} as a sum of a local squeeze operator
\begin{equation}
K(z)=\frac{1}{2\I}\Big[\hat{a}_A(z)\hat{b}^A(z)-\big(\hat{a}_A(z)\hat{b}^A(z)\big)^\dagger\Big],\label{squeezeop}
\end{equation}
 which is responsible for the conformal transformation of the fiducial area element, and the $U(1)$ generator \begin{equation}
L(z)= \frac{1}{2}\Big[\hat{a}^\dagger_A(z)\hat{a}^A(z)-\hat{b}^\dagger_A(z)\hat{b}^A(z)\Big].\label{Ugenrtr}
\end{equation}
We then have
\begin{equation}
\boldsymbol{\colon}\!C(z)\boldsymbol{\colon}=-\frac{2}{\beta^2+1}(K(z)-\beta L(z)),
\end{equation}
which is a Lorentz invariant version of the so-called linear simplicity constraints \cite{flppdspinfoam,Engle:2007uq}.

It is then straightforward to see that the reality conditions generate  the following gauge transformations, namely
\begin{align}\nonumber
\mathrm{exp}\Big({\I\int_{\mathcal{C}}\lambda\boldsymbol{\colon}\!C\boldsymbol{\colon}}\Big)\,\hat{a}^A\Big[d^2\Omega,n^\alpha\Big]&\!(z)\,\mathrm{exp}\Big(-{\I\int_{\mathcal{C}}\lambda\boldsymbol{\colon}\!C\boldsymbol{\colon}}\Big)=\\
&=\E^{-\frac{\I\beta}{\beta^2+1}\lambda(z)}\hat{a}^A\Big[\E^{\frac{2\lambda}{\beta^2+1}}d^2\Omega,n^\alpha\Big](z),\label{squeeze1}\\\nonumber
\mathrm{exp}\Big({\I\int_{\mathcal{C}}\lambda\boldsymbol{\colon}\!C\boldsymbol{\colon}}\Big)\,\hat{b}^A\Big[d^2\Omega,n^\alpha\Big]&\!(z)\,\mathrm{exp}\Big(-{\I\int_{\mathcal{C}}\lambda\boldsymbol{\colon}\!C\boldsymbol{\colon}}\Big)=\\
&=\E^{-\frac{\I\beta}{\beta^2+1}\lambda(z)}\hat{b}^A\Big[\E^{\frac{2\lambda}{\beta^2+1}}d^2\Omega,n^\alpha\Big](z),\label{squeeze2}
\end{align}
where we used the notation $\hat{a}^A[d^2\Omega,n^\alpha](z)$ to stress that the annihilation operators depend (as a functional) on the fiducial background structures $d^2\Omega$ and $n^\alpha$, and (as an ordinary function) on the points $z\in\mathcal{C}$ (see also the definition of $\hat{a}^A$ and $\hat{b}^A$ in \eref{harmop1} and \eref{harmop2} above). 

Two states in the Fock space are then said to be gauge equivalent, if there is a local gauge parameter $\lambda:\mathcal{C}\rightarrow\R$ that maps one state into the other, in other words
\begin{equation}
\Psi\sim\Psi'\Leftrightarrow\exists\lambda:\mathcal{C}\rightarrow\R:\Psi'=
\mathrm{exp}\Big({\I\int_{\mathcal{C}}\lambda\,\boldsymbol{\colon}\!C\boldsymbol{\colon}}\Big)\Psi.\label{gaugequi}
\end{equation}
In particular, any two Fock vacua that only differ by a choice for the area density $d^2\Omega$ are gauge equivalent,
\begin{equation}
\Big|0,\big\{d^2\Omega,n^\alpha\big\}\Big\rangle\sim\left|0,\big\{\E^{\frac{2\lambda}{\beta^2+1}}d^2\Omega,n^\alpha\big\}\right\rangle.\label{gaugefock}
\end{equation}
By imposing the reality conditions at the quantum level, the dependence of the Fock vacuum on the fiducial background area density $d^2\Omega$ is therefore simply washed away. Still, the boundary Fock vacuum depends on a choice for a fiducial four-normal $n^\alpha$. This dependence remains, but it is a result of having only quantised the boundary. 
Had we quantised also gravity in the bulk, we would have had to impose the glueing conditions \eref{glu1} as well. At the classical level, they are given by the constraint
\begin{align}
C_{AB}&=\frac{4\pi\I\beta G}{\beta+\I}\oepsilon^{ab}\Sigma_{ABab}-\ell_{(A}\pi_{B)}=0,\label{glucondq}
\end{align}
which links the boundary spinors $\pi_A$ and $\ell^A$ to the pull-back of the self-dual area two-form $\Sigma_{ABab}$. 
Now, $\hat{\ell}_{(A}\hat{\pi}_{B)}$ is the self-dual generator of local $SL(2,\C)$ frame rotations. For a local gauge element $\ou{\Lambda}{A}{B}:\mathcal{C}\rightarrow\mathfrak{sl}(2,\C)$, we find, in fact
\begin{align}\nonumber
\mathrm{exp}&\Big(\int_{\mathcal{C}}\I\Lambda^{AB}\hat{\pi}_A\hat{\ell}_B-\HC\Big)\,\hat{a}^A\Big[d^2\Omega,n^\alpha\Big]\!(z)\\
&\times\mathrm{exp}\Big(-\int_{\mathcal{C}}\I\Lambda^{AB}\hat{\pi}_A\hat{\ell}_B-\HC\Big)=
\ou{g}{A}{B}(\Lambda)\,\hat{a}^B\Big[d^2\Omega,\ou{g}{\alpha}{\beta}(\Lambda)n^\beta\Big](z),\label{Lorentztrafo}
\end{align}
equally for $\hat{b}^A(z)$, where $g(\Lambda)$ denotes the $SO(1,3)$ respectively $SL(2,\C)$ gauge transformation $g(\Lambda)=\mathrm{exp}(\Lambda)$. The Fock vacuum at the boundary depends parametrically on a future oriented four-vector $n^\alpha$, and any two different choices for $n^\alpha$ are related, therefore, by a Lorentz transformation that sends one vacuum into the other,
\begin{equation}
\mathrm{exp}\Big(\int_{\mathcal{C}}\I\Lambda^{AB}\hat{\pi}_A\hat{\ell}_B-\HC\Big)\Big|0,\big\{d^2\Omega,n^\alpha\big\}\Big\rangle=\left|0,\Big\{d^2\Omega,\ou{g}{\alpha}{\beta}(\Lambda)n^\beta\Big\}\right\rangle,\label{gaugefock}
\end{equation}which is a direct consequence of \eref{Lorentztrafo}.
The boundary Fock vacuum is therefore only Lorentz covariant, but not Lorentz invariant.  At the level of the Hamiltonian theory, it can be shown (see \cite{Wieland:2017zkf} for references) that the glueing conditions are the generators of simultaneous $SL(2,\C)$ gauge transformations in the bulk plus boundary. Hence we expect that local Lorentz invariance of the boundary states can be restored only by the coupling to the bulk, such that the quantum states for the bulk plus boundary geometry are entangled, $\Psi=\int d^3 n \,\Psi^{\partial{\varSigma}}_n\otimes\Psi^{\varSigma}_n$, and local $SL(2,\C)$ gauge invariance follows from the average over all possible directions of $n^\alpha$.  
\vspace{1em}


\section{Topological quantisation}\label{sec4}

In the previous section, we developed the quantisation of the gravitational boundary fields  
using a Landau representation for the boundary spinors $\ell^A$ and $\pi_A$. The goal of this section is to explain the compatibility with loop quantum gravity in the usual Ashtekar\,--\,Lewandowski representation. 

The Fock vacuum \eref{Fockvac} at the boundary depends parametrically on a choice for a fiducial area density $d^2\Omega$ and a Hermitian metric $\delta_{AA'}=\sigma_{AA'\alpha}n^\alpha$. The dependence on $d^2\Omega$ is gauged away by imposing the reality conditions \eref{realcondfock}. Two different Fock vacua that differ only by a choice for $d^2\Omega$ are gauge equivalent, and the Fock vacuum $|0,\{d^2\Omega,n^\alpha\}\rangle$ is therefore gauge equivalent
 to a totally squeezed state such as
\begin{equation}
\Psi_\emptyset=\lim_{t\rightarrow\infty}\left|0,\Big\{\E^{-\frac{2\lambda t}{\beta^2+1}}d^2\Omega,n^\alpha\Big\}\right\rangle,\label{ALvacuum}
\end{equation}
for a gauge parameter $\lambda:\mathcal{C}\rightarrow\R_>$ as in \eref{gaugefock} above. 
Such a totally squeezed state does not exist as a vector in the Hilbert space (it does not define a Cauchy sequence). Formally, it yields an eigenstate of the momentum operator $\hat{\pi}_A$ with vanishing eigenvalue (this can be seen from the definition of the annihilation operators \eref{Landauop} by sending ${}^\circ\bold{\varepsilon}_{ab}\equiv d^2\Omega\rightarrow 0$, hence  ${}^\circ\bold{\varepsilon}^{ab}\rightarrow\infty$). Using a functional Schrödinger representation, we then formally have
\begin{equation}
\forall z\in\mathcal{C}:-\I\frac{\delta}{\delta\ell^A(z)}\Psi_\emptyset[\ell^A]=-\I\frac{\delta}{\delta\bar{\ell}^{A'}(z)}\Psi_\emptyset[\ell^A]=0,\label{ALvac}
\end{equation}
hence $\Psi_\emptyset[\ell^A]=\mathrm{const}$. In loop quantum gravity, such a state  is very well known: it represents the spinorial analogue of the Ashtekar\,--\,Lewandowski vacuum \cite{Ashtekar:1993wf,Ashtekar:1994mh,LOSTtheorem} restricted to the corner. Yet in here, this state appears as just one representative of an infinite family of gauge equivalent states \eref{gaugefock}.\vspace{1em}

Let us now see how to build excited states over this vacuum and impose the reality conditions \eref{realcond} at the quantum level. The basic idea is to look at topological excitations for which the vacuum \eref{ALvacuum} is excited only over a certain number of punctures $z_1,z_2,\dots z_N\in\mathcal{C}$, such that the quantum state of the two-dimensional geometry can be described by an {$N$ body wave function}
\begin{equation}
\Psi_{f}[\ell^A]=f\big(\ell^A(z_1),\dots,\ell^A(z_N)\big),\label{wavefnctnl}
\end{equation}
in a yet unspecified $N$ particle Hilbert space $\mathcal{H}_N$. The entire boundary Hilbert space will be then given as a direct sum
\begin{equation}
\mathcal{H}=\C\oplus\mathcal{H}_1\oplus\mathcal{H}_2\oplus\dots
\end{equation}
of all $N$ particle Hilbert spaces. For the moment, the statistics is left unspecified. In particular, all punctures are thought to be distinguishable. 

At the level of the spin bundle, the $N$ body wave function $f(\ell^A_1,\dots,\ell^A_N)$ sends the $\C^2_{z_i}$ fibres\footnote{We are considering the spin bundle $\mathbb{S}(\mathcal{C},\C^2,\pi_{\mathbb{S}})$ over the corner, where each fibre $\C^2_{z}=\pi_{\mathbb{S}}^{-1}(z)$ over a point $z\in\mathcal{C}$ is homeomorphic to $\C^2$.} over the punctures into the complex numbers, hence it defines a map
\begin{equation}
f:\C^2_{z_1}\times\cdots\times\C^2_{z_N}\rightarrow\C.
\end{equation}
The actual location of the punctures $\underline{z}=(z_1,\dots,z_N)$ is gauged away by the action of \emph{small} diffeomorphisms: Consider a diffeomorphism $\varphi=\exp(\xi):\mathcal{C}\rightarrow\mathcal{C}$ at the corner that admits a horizontal lift   $\varphi_\uparrow=\exp(\xi_\uparrow):{\it\mathbb{S}}\rightarrow\mathbb{S}$ into the spin bundle $\mathbb{S}(\mathcal{C},\C^2,\pi_{\mathbb{S}})$ (the base manifold is the corner itself, the standard fibre is $\C^2$). Two states are then said to be gauge equivalent, if the bundle morphism $\varphi_\uparrow$ sends one state into the other, that is
\begin{equation}
\Psi_f\sim\Psi_{\varphi^\ast_\uparrow f},\quad\text{where:}\quad\Psi_{\varphi^\ast_\uparrow f}[\ell^A]:=\Psi_f[\varphi_\uparrow\circ\ell^A].
\end{equation}
For large gauge transformations, on the other hand, we expect that they have a non-trivial action in the quantum theory \cite{Pithis:2014uva}. This observation could be used, in fact, to determine the statistics of the $N$ particle wave function (the exchange of two punctures can always be made undone by performing a large diffeomorphism).  For the time being, we content ourselves with considering only the simplest case, where all punctures are thought to be distinguishable.\footnote{The null boundary is a three-dimensional manifold, and it seems quite plausible therefore that anyonic statistics will play an important role if the dynamics is taken into account as well, see \cite{Pithis:2014uva} for detailed thoughts about this idea.}

What is then the inner product on the $N$ particle Hilbert space? First of all, we require that the left translation along the fibres
\begin{equation}
(U_g\Psi_{f})[\ell^A]=f\Big(\ou{[g^{-1}(z_1)]}{A}{B}\ell^B(z_1),\dots,\ou{[g^{-1}(z_N)]}{A}{B}\ell^B(z_N)\Big)\label{unreprsnt}
\end{equation}
for local $SL(2,\C)$ gauge transformations $g$ be unitary. This restricts severely the functional form of $f$. The principal series of the unitary and irreducible representations of $SL(2,\C)$ are labelled and uniquely characterised by two numbers, namely by a spin $j=\Z/2$ and an additional quantum number $\rho\in\R$, and together they parametrise the two Casimir operators $\vec{L}^2-\vec{K}^2$ and $\vec{L}\cdot\vec{K}$ of the Lorentz group, see \cite{ruhl} for a detailed account. A concrete realisation of these $(\rho,j)$-representations of $SL(2,\C)$ is given by homogenous functions,
\begin{equation}
\forall \zeta\in\C-\{0\}:f_{\rho,j}\big(\zeta\ell^A\big)=\zeta^{-\I\rho+j-1}\bar{\zeta}^{-\I\rho-j-1}f_{\rho,j}\big(\ell^A\big),\label{basisstate}
\end{equation}%
where $SL(2,\C)$ acts as in \eref{unreprsnt} above. The most general $N$ body wave function \eref{wavefnctnl} can be then built from complex superpositions  of such homogenous functions, which should satisfy for all $\zeta_i\in\C-\{0\}$ and $i=1,\dots, N$ that
\begin{align}\nonumber
f_{\underline{\rho},\underline{j}}\big(\ell^A(z_1),\dots ,&\zeta\ell^A(z_i),\dots,\ell^A(z_N)\big)=\\
&=\zeta^{-\I\rho_i+j_i-1}\bar{\zeta}^{-\I\rho_i-j_i-1}\,f_{\underline{\rho},\underline{j}}\big(\ell^A(z_1),\dots ,\ell^A(z_N)\big).\label{homfunct}
\end{align}

What is then the measure with respect to which these states are normalised? The integration measure on $\C^2$
\begin{equation}
d^4\ell=\frac{1}{16}\di\ell_A\wedge\di\ell^A\wedge\di\bar{\ell}_{A'}\wedge\di\bar{\ell}^{A'},
\end{equation}
is clearly $SL(2,\C)$ invariant, but the homogenous functions are not normalisable with respect to the $L^2(\C^2,d^4\ell)$ inner product. The divergence can be removed, however, by dividing out the integration over the gauge orbits of the reality conditions \eref{realcond2}. We introduce the vector field
\begin{equation}
V_C=\frac{\I}{\beta+\I}\ell^A\frac{\partial}{\partial\ell^A}-\frac{\I}{\beta-\I}\bar\ell^{A'}\frac{\partial}{\partial\bar\ell^{A'}}.
\end{equation}
and define the three-form
\begin{equation}
d^3\mu(\ell)=V_C\hook d^4\ell=\frac{1}{8}\frac{\I}{\beta+\I}\ell_A\di\ell^A\wedge\di\bar{\ell}_{A'}\wedge\di\bar{\ell}^{A'}+\CC,
\end{equation}
where \qq{$\hook$} denotes the interior product. The inner product between two $N$ particle states is then given by the integral
\begin{equation}
\left\langle\Psi_f,\Psi_{f'}\right\rangle_{N}=\int_{\C^2_{z_1}/C}d^3\mu(\ell_1)\cdots\int_{\C^2_{z_N}/C}d^3\mu(\ell_N)\,\overline{f(\ell^A_1,\dots)}{f^\prime(\ell^A_1,\dots)},\label{innprod}
\end{equation}
where we integrate over a gauge fixing surface, such as $G(\ell_i)=\|\ell_i\|^2=\mathrm{const.}$ that intersects the gauge orbits $\ell^A_i\sim\E^{\frac{\I}{\beta+\I}\lambda(z_i)}\ell^A_i$ of the reality conditions \eref{realcond2} exactly once. The inner product is now invariant under small deformations of the gauge fixing surface, if and only if the integrant satisfies for all $i=1,\dots,N$ the constraint
\begin{equation}
\Big[\Big(\frac{\I}{\beta+\I}\ell^A_i\frac{\partial}{\partial\ell^A_i}+2\Big)+\CC\Big]\overline{f(\ell^A_1,\dots,\ell^A_N)}f'(\ell^A_1,\dots,\ell^A_N)=0.\label{gaugecond}
\end{equation}
This condition is found by deforming the gauge fixing surface (e.g.\ $G(\ell_i)=\|\ell_i\|^2=\mathrm{const.}$)  and using Stokes's theorem. Suppose now that this condition is satisfied (we will see in a moment that physical states always satisfy this condition). It then follows that the inner product is $SL(2,\C)$ gauge invariant, such that equation \eref{unreprsnt} realises a unitary representation of $SL(2,\C)$. States with different homogeneity weights are then necessarily orthogonal. 

Next, we define operators acting on these states. Consider an open neighbourhood $\mathcal{U}_z\subset\mathcal{C}$ around a point $z\in\mathcal{C}$ and define the following smeared Euler homogeneity operators, namely
\begin{align}
E_N[\mathcal{U}_{z}]\Psi[\ell^A]&:={\int_{\mathcal{U}_{z}}}\,\ell^A\big(z'\big)\frac{\delta}{\delta\ell^A(z')}\,\Psi[\ell^A],\\
\bar{E}_N[\mathcal{U}_{z}]\Psi[\ell^A]&:={\int_{\mathcal{U}_{z}}}\bar\ell^{A'}\big(z'\big)\frac{\delta}{\delta\bar\ell^{A'}(z')}\Psi[\ell^A].
\end{align}
Consider then one of our basis states, which are built from homogenous functions $f_{\underline{\rho},\underline{j}}$ as in equation \eref{homfunct} above, and define the corresponding wave functional \begin{equation}
\Psi_{f_{\underline{\rho},\underline{j}}}[\ell^A]=f_{\underline{\rho},\underline{j}}\big(\ell^A(z_1),\dots,\ell^A(z_N)\big).\label{punctstate}
\end{equation}
A short moment of reflection reveals that any such state is an eigenvector of the Euler operators with eigenvalues given by
\begin{align}\label{Euler1}
E_N[\mathcal{U}_{z}]\Psi_{f_{\underline{\rho},\underline{j}}}
&=\sum_{i=1}^N\chi_{\mathcal{U}_{z}}(z_i)\big(-\I\rho_i+j_i-1\big)\Psi_{f_{\underline{\rho},\underline{j}}},\\
\bar{E}_N[\mathcal{U}_{z}]\Psi_{f_{\underline{\rho},\underline{j}}}
&=\sum_{i=1}^N\chi_{\mathcal{U}_{z}}(z_i)\big(-\I\rho_i-j_i-1)\Psi_{f_{\underline{\rho},\underline{j}}},
\end{align}
where $\chi_{\mathcal{U}_z}(z')$ denotes the characteristic function of $\mathcal{U}_z\subset\mathcal{C}$. If \eref{gaugecond} is satisfied, the inner product is $SL(2,\C)$ invariant. States of different homogeneity weights are then orthogonal and the adjoint operators must satisfy, therefore,
\begin{align}\label{Euler2}
E^\dagger_N[\mathcal{U}_{z}]\Psi_{f_{\underline{\rho},\underline{j}}}
&=\sum_{i=1}^N\chi_{\mathcal{U}_{z}}(z_i)\big(\I\rho_i+ j_i-1\big)\Psi_{f_{\underline{\rho},\underline{j}}},\\
\bar{E}^\dagger_N[\mathcal{U}_{z}]\Psi_{f_{\underline{\rho},\underline{j}}}
&=\sum_{i=1}^N\chi_{\mathcal{U}_{z}}(z_i)\big(\I\rho_i-j_i-1)\Psi_{f_{\underline{\rho},\underline{j}}}.
\end{align}

Next we have to quantise the reality conditions \eref{realcond2} and find their kernel in the state space spanned by the homogenous wave functions \eref{punctstate}. At the classical level, the reality conditions imply that for any open neighbourhood $\mathcal{U}_z\subset{\mathcal{C}}$ around any point $z\in\mathcal{C}$ the constraint 
\begin{equation}
\frac{\I}{\beta+\I}\int_{\mathcal{U}_z}\pi_A\ell^A+\CC=0
\end{equation}
is satisfied. Choosing a symmetric ordering, we define the operator
\begin{equation}
\boldsymbol{\colon}\!{\int_{\mathcal{U}_z}\pi_A\ell^A}\boldsymbol{\colon}=\frac{1}{2\I}\big(E_N[\mathcal{U}_z]-\bar{E}^\dagger_N[\mathcal{U}_z]\big).
\end{equation}
Physical states $\Psi_{\text{phys}}$ are now given by those wave functionals that are annihilated by the reality conditions, hence
\begin{equation}
\Big[\frac{\I}{\beta+\I}\boldsymbol{\colon}\!{\int_{\mathcal{U}_z}\pi_A\ell^A}\boldsymbol{\colon}+\HC\Big]\Psi_{\mathrm{phys}}=0.\label{realcondq}
\end{equation}
It is now immediate to impose the reality conditions, and identify their solution space. The eigenvalues of the Euler homogeneity operators are given in \eref{Euler1} and \eref{Euler2}, such that the only allowed values for the quantum numbers $\rho_i$ and $j_i$ must satisfy the relation
 \begin{equation}
\frac{\I}{\beta+\I}(\rho_i+\I\,j_i)+\CC=0\Leftrightarrow\rho_i=\beta j_i.\label{realcondspectrm}
\end{equation}

In defining the inner product \eref{innprod}, we mentioned that the integrals are independent of the gauge fixing only if the integrant satisfies the constraint \eref{gaugecond}. We can now verify this condition: any $N$ body physical state can be written as a superposition of homogenous functions
\begin{equation}
\Psi_{\mathrm{phys}}[\ell^A]=\sum_{j_1\dots j_N} c^{j_1\dots j_N} f_{(\beta j_1\dots\beta j_N),(j_1\dots j_N)}\big(\ell^A(z_1),\dots,\ell^A(z_N)\big)
\end{equation}
that all satisfy for all $i=1,\dots, N$ the reality conditions $\rho_i=\beta j_i$. From there, it is easy to see that \eref{gaugecond} is satisfied, such that the inner product \eref{innprod} between physical states is indeed independent of the gauge fixing. The $N$ particle Hilbert space is then given by the Cauchy completion (with respect to the norm induced by the inner product \eref{innprod}) of the complex span of all such normalisable states $\Psi_{\mathrm{phys}}$.
\vspace{1em}

Having imposed the reality conditions at the quantum level, we can now introduce the area operator and compare its spectrum with what we found in the last section, see \eref{arspec}. In the classical theory, the area \eref{Ardef} of a neighbourhood $\mathcal{U}_z$ on $\mathcal{C}$ is given by the integral
\begin{equation}
\mathrm{Ar}[\mathcal{U}_z]=-\I\int_{\mathcal{U}_z}\bold{\eta}_A\ell^A.
\end{equation}
Choosing a normal ordering, we can now quantise this operator simply by saying
\begin{equation}
\boldsymbol{\colon}\!{\mathrm{Ar}}[\mathcal{U}_z]\boldsymbol{\colon}=-8\pi G\frac{\beta}{\beta+\I}E[\mathcal{U}_z],
\end{equation}
where $E[\mathcal{U}_z]$ is the Euler homogeneity operator \eref{Euler1}. Consider then a homogenous function $f_{\underline{j}}\equiv f_{\underline{\rho},\underline{j}}$ such that the reality conditions \eref{realcondspectrm} are satisfied: $\rho_i=\beta j_i$ for all $i=1,\dots,N$. The corresponding wave functional $\Psi_{f_{\underline{j}}}$ is an eigenstate of the area operator, with eigenvalues given by
\begin{equation}
\boldsymbol{\colon}\!{\mathrm{Ar}}[\mathcal{U}_z]\boldsymbol{\colon}\Psi_{f_{\underline{j}}}=8\pi\,\beta G\,\sum_{i=1}^N\chi_{\mathcal{U}_z}(z_i)j_i\;\Psi_{f_{\underline{j}}}.\label{arspec2}
\end{equation}
The possible eigenvalues of the area of the entire corner are then simply given by
\begin{equation}
a_n=4\pi\,\beta G\,n,\quad n\in\Z/2,
\end{equation}
which agrees with the bosonic quantisation that we introduced earlier \eref{arspec}. Hence, we arrive at the same conclusion as before. In quantum gravity area is quantised.

\vspace{1em}

Before concluding, one last remark. The interpretation of $\Psi_f$ as a wave \emph{function} for $N$ particles is actually quite appropriate. Suppose, for a moment, that the classical metric in the neighbourhood of the corner is flat, and that the corner itself is a round two-sphere. The null vector $\ell^\alpha$ that shines out of this sphere is the square $\ell^\alpha=-\uo{\sigma}{AA'}{\alpha}\ell^A\bar{\ell}^{A'}$ of the spinor
\begin{equation}
\ell^A(\vartheta,\varphi)=\begin{pmatrix}
\ell^0(\vartheta,\varphi)\\
\ell^1(\vartheta,\varphi)
\end{pmatrix}
=
\begin{pmatrix}
\mathrm{cos}\frac{\vartheta}{2}\\
\E^{\I\varphi}\mathrm{sin}\frac{\vartheta}{2}
\end{pmatrix}.
\end{equation}
Using a stereographic projection, we can then use this spinor itself as a coordinate on the sphere, such that a point is marked by the ratio 
\begin{equation}
\frac{\ell^1(z)}{\ell^0(z)}=z=\E^{\I\varphi}\mathrm{tan}\frac{\vartheta}{2}\label{complexcoord}
\end{equation}
of the spin up and down components. The $N$ body homogenous wave functions \eref{punctstate} can be then written as a product of a universal prefactor times an $N$ body wave function $f(z_1,\dots,z_N)$, which only depends on the complex coordinates \eref{complexcoord} labelling the locations $z_1,\dots,z_N$ of the punctures on the sphere. The most general state in the $N$ body Hilbert space can be then written in the following form,
\begin{align}\nonumber
\Psi_{{f}}[\ell^A]=\sum_{j_1,\dots,j_N}c_{j_1\dots j_N}\prod_{i=1}^N\big(\ell^0(z_i)\big)^{-\I j_i(\beta+\I)-1}&\big(\bar{\ell}^{\,0'}(z_i)\big)^{-\I j_i(\beta-\I)-1}\\
&\times f_{j_1\dots j_N}(z_1,\dots,z_N),
\end{align}
for complex coefficients $\{c_{j_1\dots j_N}\}$ and $N$ body wave functions $f_{j_1\dots }(z_1,\dots)$ on the fiducial round sphere.

\section{Summary and conclusion}
Let us summarise. First of all (\hyperref[sec2]{section 2}), we studied the boundary symplectic structure at a two-dimensional cross-section of a null surface (see \hyperref[fig2]{figure 2}). The canonically conjugate variables at the boundary consist of a spinor $\ell^A$ and a spinor-valued surface density $\pi_A$. We then showed that the area of the cross section can be written as a surface integral of the Lorentz invariant contraction (i.e.\ the helicity) of the spinors,
\begin{equation}
\mathrm{Ar}[\mathcal{C}]=4\pi\I\,\beta G\int_{\mathcal{C}}\big(\pi_A\ell^A-\CC).\label{areaop}
\end{equation}
To be compatible with a real and Lorentzian metric, the spinors have to satisfy certain constraints, namely the reality conditions \eref{realcond2}.
\vspace{1em}

Next (\hyperref[sec3]{section 3}), we built two pairs of harmonic oscillators out of the spinors. This required additional fiducial background structures, namely a two-dimensional surface density $d^2\Omega$ and a Hermitian metric $\delta_{AA'}=\sigma_{AA'\alpha}n^\alpha$. We then quantised the oscillators using a bosonic representation obtaining a Fock vacuum $|0,\{d^2\Omega,n^\alpha\rangle$, which depends parametrically on the fiducial background structures.  The oriented area \eref{areaop} of the cross section turned into the difference of two number operators, with no dependence on the fiducial background structures. In quantum theory, the area  becomes quantised (the integral \eref{areaop} is essentially the generator of global $U(1)$ transformation of the spinors), and the spectrum of the cross sectional area is equidistant.\footnote{Such area spectra also appear in other approaches, such as the semi-classic quantisation of black holes due to Bekenstein\,--\,Mukhanov \cite{BekensteinMukhanovEffect}, and non-commutative geometry \cite{AmelinoCamelia:2008wq,Chamseddine:2014nxa}.} The possible eigenvalues have infinite degeneracy and are all multiples of the fundamental loop gravity area gap $
a_o={8\pi\beta\,\hbar G}/{c^3}$.
\vspace{1em}

Finally (\hyperref[sec4]{section 4}), we studied how the Fock representation for the boundary fields fits together with loop quantum gravity in the usual spin network representation. We imposed the reality conditions \eref{realcond2} at the quantum level, and saw that two different Fock vacua $|0,\{d^2\Omega_1,n^\alpha\rangle$ and $|0,\{d^2\Omega_2,n^\alpha\rangle$ that differ only by a choice for the fiducial area element are gauge equivalent (the reality conditions are the sum of a squeeze operator \eref{squeezeop} and an $U(1)$ generator \eref{Ugenrtr}, the squeeze operator generating conformal transformations of the fiducial area density $d^2\Omega$). Imposing the reality conditions amounts therefore to consider only gauge equivalence classes\footnote{There are no normalisable states in the Fock space that would lie in the kernel of the reality conditions \eref{realcondfock}. The Fock space that we introduced in \hyperref[sec3]{section 3} is therefore only kinematical, physical states represent distributions (elements of the algebraic dual of the Fock space). Yet by duality, the area spectrum can already be inferred from the states in the Fock space alone.} of states that are related by the gauge transformations (\ref{squeeze1}, \ref{squeeze1}). Thus, we washed away the dependence of the Fock vacuum on the fiducial background area density $d^2\Omega$.  The Fock vacuum for any given choice of $d^2\Omega$ is therefore gauge equivalent to a totally squeezed state such as \eref{ALvacuum}. Such a totally squeezed state is well known from loop quantum gravity, where  it represents the analogue of the Ashtekar\,--\,Lewandowski vacuum at a two-dimensional surface. Finally, we introduced topological excitations over this vacuum, such that the boundary spinors are excited only over a finite number of punctures $z_1,\dots,z_N\in\mathcal{C}$. Every such puncture carries a unitary representation of the Lorentz group, which are classified by quantum numbers $\rho\in\R$ and $j\in\Z/2$.  We then quantised the reality conditions and imposed them at the quantum level: Only those representations contribute for which $\rho=\beta j$, which is the same kind of constraint that appears in the definition of the loop gravity transition amplitudes \cite{flppdspinfoam,Engle:2007uq}. Once the reality conditions are imposed, the spectrum of the area operator is therefore discrete, and it matches the one that we derived using the Fock representation in the continuum, see \eref{arspec} and \eref{arspec2}.

\vspace{1em}

In summary, our quantisation of the gravitational boundary modes answers and addresses some long standing and well-founded doubts and reservations  against loop gravity, see \cite{outside}. It was often remarked that the derivation of the area spectrum relies on using $SU(2)$ gauge connection variables, whereas the geometrically relevant choice for gravity seems to be rather $SL(2,\C)$. Our calculation is manifestly Lorentz invariant (this becomes particularly clear in \hyperref[sec4]{section 4}), and no gauge fixing to compact gauge groups is ever required. Another critique came from the usage of spin network functions. It was argued that by working with spin network functions the discreteness was introduced right from the onset. Yet, in our framework all fields are continuous, and no discretisation was ever introduced. In addition, we have quantised the boundary variables on a null surface, such that the discreteness of area in quantum gravity seems to be indeed compatible with both local Lorentz invariance and the universal causal structure of the light cone.

\paragraph{Acknowledgments} This research was supported in part by Perimeter Institute for Theoretical Physics. 
Research at Perimeter Institute is supported by the Government of Canada through the Department of Innovation, Science and Economic Development and by the Province of Ontario through the Ministry of Research and Innovation.

\providecommand{\href}[2]{#2}\begingroup\raggedright\endgroup


\begin{thebibliography}{10}

\bibitem{Ashtekar:2000eq}
A.~Ashtekar, J.~C. Baez, and K.~Krasnov, ``{Quantum geometry of isolated
  horizons and black hole entropy},'' {\em Adv. Theor. Math. Phys.} {\bf 4}
  (2000) 1--94,
\href{http://arXiv.org/abs/gr-qc/0005126}{{\tt arXiv:gr-qc/0005126}}.

\bibitem{PhysRevD.82.044050}
J.~Engle, K.~Noui, A.~Perez, and D.~Pranzetti, ``Black hole entropy from the
  $SU(2)$-invariant formulation of type I isolated horizons,'' {\em Phys. Rev.
  D} {\bf 82} (2010), no.~4, 044050, \href{http://arXiv.org/abs/1006.0634}{{\tt
  arXiv:1006.0634}}.

\bibitem{Sahlmann:2011aa}
H.~Sahlmann, ``Black hole horizons from within loop quantum gravity,'' {\em
  Phys. Rev. D} {\bf 84} (2011) 044049,
  \href{http://arXiv.org/abs/1104.4691}{{\tt arXiv:1104.4691}}.

\bibitem{wieland:nulldefects}
W.~Wieland, ``Discrete gravity as a topological field theory with light-like
  curvature defects,'' {\em JHEP} {\bf 5} (2017) 1--43,
  \href{http://arXiv.org/abs/1611.02784}{{\tt arXiv:1611.02784}}.

\bibitem{Wieland:2017zkf}
W.~Wieland, ``{New boundary variables for classical and quantum gravity on a
  null surface},''
\href{http://arXiv.org/abs/1704.07391}{{\tt arXiv:1704.07391}}.

\bibitem{selfdualtwo}
R.~Capovilla, T.~Jacobson, J.~Dell, and L.~Mason, ``{Selfdual two forms and
  gravity},'' {\em Class. Quant. Grav.} {\bf 8} (1991)
41--57.

\bibitem{AshtekarLewandowskiArea}
A.~Ashtekar and J.~Lewandowski, ``{Quantum theory of geometry I.: Area
  operators},'' {\em Class. Quant. Grav.} {\bf 14} (1997) A55--A82,
\href{http://arXiv.org/abs/gr-qc/9602046}{{\tt arXiv:gr-qc/9602046}}.

\bibitem{Rovelliarea}
C.~Rovelli and L.~Smolin, ``Discreteness of area and volume in quantum
  gravity,'' {\em Nuclear Physics B} {\bf 442} (1995), no.~3, 593--619,
  \href{http://arXiv.org/abs/gr-qc/9411005}{{\tt arXiv:gr-qc/9411005}}.

\bibitem{Wieland:2016dbc}
W.~Wieland, ``{Quasi-local gravitational angular momentum and centre of mass
  from generalised Witten equations},'' {\em Gen. Rel. Grav.} {\bf 49} (2017),
  no.~3, 38,
\href{http://arXiv.org/abs/1604.07428}{{\tt arXiv:1604.07428}}.

\bibitem{twist}
L.~Freidel and S.~Speziale, ``{Twistors to twisted geometries},'' {\em Phys.
  Rev. D} {\bf 82} (Oct, 2010) 084041,
  \href{http://arXiv.org/abs/1006.0199}{{\tt arXiv:1006.0199}}.

\bibitem{komplexspinors}
W.~Wieland, ``{Twistorial phase space for complex Ashtekar variables},'' {\em
  {Class. Quant. Grav.}} {\bf 29} (2011) 045007,
  \href{http://arXiv.org/abs/1107.5002}{{\tt arXiv:1107.5002}}.

\bibitem{Bianchi:2016hmk}
E.~Bianchi, J.~Guglielmon, L.~Hackl, and N.~Yokomizo, ``{Loop expansion and the
  bosonic representation of loop quantum gravity},'' {\em Phys. Rev. D} {\bf
  94} (2016), no.~8, 086009,
\href{http://arXiv.org/abs/1609.02219}{{\tt arXiv:1609.02219}}.

\bibitem{Donnelly:2016auv}
W.~Donnelly and L.~Freidel, ``{Local subsystems in gauge theory and gravity},''
  {\em JHEP} {\bf 9} (2016) 102,
\href{http://arXiv.org/abs/1601.04744}{{\tt arXiv:1601.04744}}.

\bibitem{Freidel:2015gpa}
L.~Freidel and A.~Perez, ``{Quantum gravity at the corner},''
\href{http://arXiv.org/abs/1507.02573}{{\tt arXiv:1507.02573}}.

\bibitem{Geiller:2017xad}
M.~Geiller, ``{Edge modes and corner ambiguities in 3d Chern-Simons theory and
  gravity},''
\href{http://arXiv.org/abs/1703.04748}{{\tt arXiv:1703.04748}}.

\bibitem{Dittrich:2014wpa}
B.~Dittrich and M.~Geiller, ``{A new vacuum for Loop Quantum Gravity},'' {\em
  Class. Quant. Grav.} {\bf 32} (2015), no.~11, 112001,
\href{http://arXiv.org/abs/1401.6441}{{\tt arXiv:1401.6441}}.

\bibitem{Bahr:2015bra}
B.~Bahr, B.~Dittrich, and M.~Geiller, ``{A new realization of quantum
  geometry},''
\href{http://arXiv.org/abs/1506.08571}{{\tt arXiv:1506.08571}}.

\bibitem{partobs}
C.~Rovelli, ``{Partial observables},'' {\em Phys. Rev. D} {\bf 65} (June, 2002)
  124013, \href{http://arXiv.org/abs/gr-qc/0110035v3}{{\tt
  arXiv:gr-qc/0110035v3}}.

\bibitem{flppdspinfoam}
J.~Engle, R.~Pereira, and C.~Rovelli, ``Flipped spinfoam vertex and loop
  gravity,'' {\em Nucl. Phys. B} {\bf 798} (2008) 251--290,
  \href{http://arXiv.org/abs/0708.1236v1}{{\tt arXiv:0708.1236v1}}.

\bibitem{Engle:2007uq}
J.~Engle, R.~Pereira, and C.~Rovelli, ``{The Loop-quantum-gravity
  vertex-amplitude},'' {\em Phys. Rev. Lett.} {\bf 99} (2007) 161301,
\href{http://arXiv.org/abs/0705.2388}{{\tt arXiv:0705.2388}}.

\bibitem{Ashtekar:1993wf}
A.~Ashtekar and J.~Lewandowski, ``{Representation theory of analytic holonomy
  C* algebras},'' in {\em Knots and Quantum Gravity}, J.Baez, ed.
\newblock Oxford University Press, 1993.
\newblock
\href{http://arXiv.org/abs/gr-qc/9311010}{{\tt arXiv:gr-qc/9311010}}.
\newblock

\bibitem{Ashtekar:1994mh}
A.~Ashtekar and J.~Lewandowski, ``{Projective techniques and functional
  integration for gauge theories},'' {\em J. Math. Phys.} {\bf 36} (1995)
  2170--2191,
\href{http://arXiv.org/abs/gr-qc/9411046}{{\tt arXiv:gr-qc/9411046}}.

\bibitem{LOSTtheorem}
J.~Lewandowski, A.~Okolow, H.~Sahlmann, and T.~Thiemann, ``{Uniqueness of
  diffeomorphism invariant states on holonomy-flux algebras},'' {\em Commun.
  Math. Phys.} {\bf 267} (2006) 703--733,
\href{http://arXiv.org/abs/gr-qc/0504147}{{\tt arXiv:gr-qc/0504147}}.

\bibitem{Pithis:2014uva}
A.~G.~A. Pithis and H.-C. Ruiz~Euler, ``{Anyonic statistics and large horizon
  diffeomorphisms for Loop Quantum Gravity Black Holes},'' {\em Phys. Rev. D}
  {\bf 91} (2015) 064053,
\href{http://arXiv.org/abs/1402.2274}{{\tt arXiv:1402.2274}}.

\bibitem{ruhl}
W.~R{\"u}hl, {\em {The Lorentz Group and Harmonic Analysis}}.
\newblock W. A. Benjamin, New York, 1970.

\bibitem{BekensteinMukhanovEffect}
J.~D. Bekenstein and V.~F. Mukhanov, ``{Spectroscopy of the quantum black
  hole},'' {\em Phys. Lett. B} {\bf 360} (1995) 7--12,
\href{http://arXiv.org/abs/gr-qc/9505012}{{\tt arXiv:gr-qc/9505012}}.

\bibitem{AmelinoCamelia:2008wq}
G.~Amelino-Camelia, G.~Gubitosi, and F.~Mercati, ``{Discreteness of area in
  noncommutative space},'' {\em Phys. Lett. B} {\bf 676} (2009) 180--183,
\href{http://arXiv.org/abs/0812.3663}{{\tt arXiv:0812.3663}}.

\bibitem{Chamseddine:2014nxa}
A.~H. Chamseddine, A.~Connes, and V.~Mukhanov, ``{Quanta of Geometry:
  Noncommutative Aspects},'' {\em Phys. Rev. Lett.} {\bf 114} (2015), no.~9,
  091302,
\href{http://arXiv.org/abs/1409.2471}{{\tt arXiv:1409.2471}}.

\bibitem{outside}
H.~Nicolai, K.~Peeters, and M.~Zamaklar, ``{Loop quantum gravity: an outside
  view},'' {\em Class. Quantum Grav.} {\bf 22} (2005), no.~10, R193--R347,
  \href{http://arXiv.org/abs/hep-th/0501114}{{\tt arXiv:hep-th/0501114}}.

\end{thebibliography}
\end{document}